\documentstyle[12pt,titlepage,twoside,drop,epsf]{thes}

\begin{document}
\pagenumbering{roman}
\title{\bf\Huge Effective Medium Tight-Binding Theory}
\author{\vspace{7.5cm}\\
\includegraphics{plots/meltslab.ps}
\\{\normalsize\rm Ph.D. Thesis August 1994}\\
{\large Kurt Stokbro}\\
{\normalsize\rm Physics  Department}\\
{\normalsize\rm Technical University of Denmark}\\
{\normalsize\rm DK-2800 Lyngby, Denmark }}
\date{}
\maketitle
\mbox{}\thispagestyle{empty}\newpage
\thispagestyle{empty}
\vspace*{17.5cm}
\hspace*{3.9cm}
\begin{minipage}{9.3cm}
``In the cases where the laws of mathematics describe
reality they are not exact, and in the cases they are exact they do
not describe reality.''
\begin{flushright}
{\it A. Einstein}
\end{flushright}
\end{minipage}
\newpage
\mbox{}\thispagestyle{empty}\newpage
\thispagestyle{empty}
\vspace*{-1cm}
{\center\bf\large Effective Medium Tight-Binding Theory \\}
{\center  Kurt Stokbro \\}
\vspace{-5mm}
{\center Physics  Department  \\}
\vspace{-5mm}
 {\center Technical University of Denmark \\}
\vspace{-5mm}
{\center  DK-2800 Lyngby, Denmark \\}
{\center{\bf Abstract} \\}
\mbox{}\newline
\begin{minipage}{14cm}
\footnotesize
\hspace*{6mm}
In this thesis a new model for calculating the total energy of atomic and
solid systems is presented. The
model is tested on both aluminum and silicon systems, and
has been implemented in a molecular dynamics program. This program is
used to study the properties of the
 Si(100) surface and dislocation dynamics in the diamond structure. For
the Si(100) surface the main result is the finding of an incomplete
melted state at $1550^0 K$. The studied dislocation is the $90^0$ partial and
it is found that the barriers for kink formation and migration
are in good accordance with experimental observations and related to the energy
difference between a symmetric and  asymmetric reconstruction of the
dislocation core.
The model for the total energy is derived from
non-selfconsistent density-functional theory and uses an electronic density
given by a superposition of optimized atom-based densities.
The general idea for calculating the total energy of a given system
is  to relate a reference system to each atom, an effective medium for
which the total energy is known and then only calculate the energy
difference between the two systems. The reference system is chosen such
that the charge neutral sphere around an atom has the same radius in
both systems, and with this choice of reference system the energy difference
between the
two systems become small, and can therefore be calculated approximately.
{}From the optimized densities a pair potential is constructed which
describe the electro-static and exchange-correlation part of the energy
difference. The remaining part, the one-electron difference, is
calculated with an LMTO tight-binding model with the potential parameters
determined from the effective potential of the reference system.
\footnotetext{{\bf Descriptors}: Total-energy methods, density-functional
theory,
LMTO tight-binding, molecular dynamics, silicon (100) surface, dislocations. }
\end{minipage}
\newpage
\mbox{}\thispagestyle{empty}\newpage
\thispagestyle{empty}
\vspace*{-1cm}
{\center\bf\large Effektiv  medium t\ae tbindingsteori \\}
{\center  Kurt Stokbro \\}
\vspace{-5mm}
{\center Fysisk Institut  \\}
\vspace{-5mm}
 {\center Danmarks Tekniske Universitet\\}
\vspace{-5mm}
{\center  DK-2800 Lyngby, Danmark \\}
{\center{\bf Resum\'{e}} \\}
\mbox{}\newline
\begin{minipage}{14cm}
\footnotesize
\hspace*{6mm}
I denne afhandling pr\ae senteres en ny model for beregning af
totalenergier af molekyler og faste stoffer. Modellen er blevet testet for
systemer best\aa ende af  Aluminium og Silicium, og er implementeret i
et molekyl\ae rt dynamikprogram. Dette program er anvendt til at studere
egenskaberne af Silicium (100) overfladen, samt dynamikken af
dislokationer i diamantstrukturen. Simulationerne viser at
 Silicium (100) overfladen laver ufuldst\ae ndig overfladesmeltning n\aa r
temperaturen n\ae rmer sig den krystalinske smeltetemperatur. Den studerede
dislokation er en partiel kantdislokation i diamantstrukturen, og
barrierne for bev\ae gelse af denne dislokation er beregnet. Disse er i god
overensstemmelse med eksperimentielle m\aa linger, og kan relateres til
rekonstruktionen af dislokationskernen.

Modellen for totalenergien er baseret p\aa \ en
r\ae kke approximationer til t\ae thedsfunktional teori, og resultatet
 er en totalenergifunktion som er beregningsm\ae ssigt
simplere, men samtidigt giver p\aa lidelige totalenergier.
Konstruktionen bygger p\aa \ Effektiv medium Teori, idet
totalenergien beregnes ved at relatere hvert atom til et reference
system, hvor atomerne har
lignende kemiske egenskaber. Idet  totalenergien af reference systemet
er kendt, er det
 kun energiforskellen mellem de to systemer der skal beregnes.
Referencesystemet er valgt s\aa ledes at en ladningsneutral kugle
omkring et atom, har samme radius i begge systemer.
Med dette valg vil de to systemer have ensartede kemiske egenskaber,
hvilket bevirker at energiforskellen mellem systemerne er lille og
den kan derfor beregnes approksimativt. Det er nu muligt at beregne det
elektrostatiske og exchangekorrelations bidraget til totalenergien
ved hj\ae lp af et parpotential. Den sidste del, en-elektronbidraget,
beregnes ved hj\ae lp af en LMTO t\ae tbindings model, hvor potential
parametrene konstrueres udfra potentialet  i referencesystemet.
\end{minipage}
\newpage
\mbox{}\thispagestyle{empty}\newpage
\chapter*{Preface}
This is a thesis for the Danish Ph.D. degree in physics, and covers my
research in approximate total-energy methods
at the Danish Technical University in the period 1 February 1991 -- 1 august
1994
under supervision of Professor Karsten Wedel Jacobsen and Professor Jens Kehlet
N\o
rskov. The research has been published in 5 articles and  the aim of
the presentation is to make a self-contained review which adjoin the
topics  covered in these articles. However, I have tried to select unpublished
material for the presentation such that the reader may gain by reading
both the thesis and the articles.

I met for the first time K.W. Jacobsen a Wednesday in January 1991, and his
enthusiasm was
so catching that I started on a Ph.D. program the week after. I have
never regreted this fast decision, he has been a very competent
supervisor who has given good advice and encouraged me when the road
seemed unpassable. After the first year I began to work with J. K. N\o
rskov and with time he has been almost as valuable to me as K.W. Jacobsen,
such that today they both  stand as my supervisors. Most importantly,
they have been good examples and have given me a joy by making
physics, such that I have decided to continue in this field.
It has also been a pleasure to work closely with Nithaya
Chetty(now at Brookhaven N.Y.), and many of the ideas in this thesis,
especially those related
to the optimized densities, are due to him.

I have found it inspiring to be in a large group where everybody works in
related areas; unfortunately this is a rare scenario in Danish
theoretical physics. I admire J.K. N\o rskov for his good leadership,
the staff complement each other very well and there are good computer
facilities. I thank all members of the group who have all been very
helpful to me, especially O. H. Nielsen and J. Sch\"{o}tz with the computers,
B. Hammer
and P. Kratzer with Car-Parrinello timing results, and A. Christensen
with the Aluminium calculations.  I thank Hans Skriver for giving me his LMTO
code and
many valuable discussions concerning the method. For the molecular-dynamics
simulations I have used a simulation package written by
Per Stoltze, and I thank him for  many valuable discussions concerning
simulation techniques and surface  melting.

In February
1994 I visited professor Kai-Ming Ho's group at Ames
Laboratory, Iowa State University, and I thank all the people for
giving me three exiting weeks, despite the dreadful weather. The
collaboration with David Deaven has been very fruitful, and together we
managed to implement the EMTB scheme in a molecular-dynamics program
during the stay. It has also been a pleasure working together with Lars
Hansen on the dislocation dynamics, and I thank both  David Deaven and
Lars Hansen for proof reading the manuscript.

For the work in this thesis I have used computer codes  provided
by several people.   I acknowledge K. Kunc, O.H. Nielsen, R.J. Needs, and R.M.
Martin
whose solid state programs I have used, and  E.L. Shirley who developed
the pseudo-potential routines. I also thank C. S. Fadley for letting
me use his
program for single-scattering cluster calculation, and J. Fraxedas for
learning me how to use it.
The first three years of my study has been financed by the Danish
Research Councils and the last half year by the Center for Atomic-Scale
Physics (CAMP), which is sponsored by the Danish National Research
Foundation.

\vspace{1cm}

\noindent
{\em Lyngby} August 1994 \hfill Kurt Stokbro

\
\setcounter{page}{1}\thispagestyle{empty}
\newpage
This thesis is based on the following papers:
\newline\thispagestyle{empty}
\vspace{1cm}
\begin{tabbing}
III.XX \= \kill
{\bf I.}
\> {\em Simple Model of Stacking-Fault Energies} \\
\> K. Stokbro and K. W. Jacobsen\\
\> Phys. Rev. B {\bf 47}, 4916 (1993)\\
\\
\\
{\bf II.}
\> {\em An ab initio Potential for Solids} \\
\> N. Chetty, K. Stokbro, K. W. Jacobsen, and J. K. N\o rskov \\
\> Phys. Rev. B {\bf 46}, 3798 (1992) \\
\\
\\
{\bf III.}
\> {\em Ab initio  Effective-Medium Theory for Al} \\
\> K. Stokbro, N. Chetty,  K. W. Jacobsen, and J. K. N\o rskov \\
\>Proceedings from the $15^{th}$ Taniguchi symposium, 15 (Springer~1993)  \\
\\
\\
{\bf IV.}
\> {\em Construction of Transferable Spherically-averaged Electron
Potentials} \\
\> K. Stokbro, N. Chetty, K. W. Jacobsen, and J. K. N\o rskov \\
\>  J. Phys. Condens Matter {\bf 6}, 5415 (1994) \\
\\
\\
{\bf V.}
\> {\em An Effective-Medium Tight-Binding Model for Silicon } \\
\> K. Stokbro, N. Chetty, K. W. Jacobsen, and J. K. N\o rskov \\
\>  To appear in Phys.~Rev~B \\
\\
\\
\end{tabbing}
\newpage
\mbox{}\thispagestyle{empty}\newpage
\tableofcontents \thispagestyle{empty}\newpage
\pagenumbering{arabic}
\setcounter{page}{1}
\include{intro}
\include{energyfu}
\include{potapp}
\include{param}
\include{moldyn}
\include{appli}
\include{conclu}
\mbox{}\thispagestyle{empty}\newpage
\bibliographystyle{thesref}
\footnotesize
\bibliography{reflist}
\newpage
\end{document}